\documentclass[twocolumn,prl, showpacs]{revtex4}
\usepackage{graphics,graphicx}
\begin{document}
\title{Electronic nature of the lock-in magnetic transition in Ce$X$Al$_{4}$Si$_{2}$}
\author{J.~Gunasekera$^{1}$}
\author{L. ~Harriger$^{2}$}
\author{A.~Dahal$^{1}$}
\author{A.~Maurya$^{3}$}
\author{T. ~Heitmann$^{4}$}
\author{S. ~Disseler$^{2}$}
\author{A.~Thamizhavel$^{3}$}
\author{S.~Dhar$^{3}$}
\author{D. K.~Singh$^{1,*}$}
\affiliation{$^{1}$Department of Physics and Astronomy, University of Missouri, Columbia, MO 65211}
\affiliation{$^{2}$NIST Center for Neutron Research, Gaithersburg, MD, 20899}
\affiliation{$^{3}$Tata Institute of Fundamental Research, Mumbia, India}
\affiliation{$^{4}$University of Missouri Research Reactor, University of Missouri, Columbia, MO 65211}
\affiliation{$^{*}$email: singhdk@missouri.edu}

\begin{abstract}

We have investigated the underlying magnetism in newly discovered single crystal Kondo lattices Ce$X$Al$_{4}$Si$_{2}$, where $X$ = Rh, Ir. We show that the compound undergoes an incommensurate-to-commensurate magnetic transition at $T_c$ = 9.19 K (10.75 K in Ir). The spin correlation in the incommensurate phase is described by a spin density wave configuration of Ce-ions, which locks-in to the long-range antiferromagnetic order at $T$ = $T_c$. The qualitative analysis of the experimental data suggests the role of the Fermi surface nesting, instead of the lattice distortion causing the Umklapp correction or the soliton propagation, as the primary mechanism behind this phenomenon.  
 
\end{abstract}

 \maketitle

Strongly correlated electrons systems provide a fertile research avenue that encompasses a host of electronic phenomena, such as quantum critical behavior, unconventional superconductivity, multiferroic and unusual electronic properties associated with the reconstruction of the Fermi surface.\cite{Varma,Sachdev,Fisk,Saxena,Mathur,Aronson,Watanabe} Among the strongly correlated electrons family, heavy electron compounds are of special interests.\cite{Stewart,Si,Coleman} Many of these materials exhibit an interplay between magnetism and unconventional superconductivity where novel magnetic quantum critical effect is found to play the key role.\cite{Broun,Schroder} The heuristic quantum critical phenomenon is often accompanied by a change in the Fermi surface properties.\cite{Steglich,Si} Another unusual magnetic phenomenon, which may be arising due to the change in the electronic properties in a magnetic material, is associated to magnetic phase transition between commensurate and incommensurate order as a function of temperature; often referred to as the “lock-in magnetic transition”.\cite{Fawcett} Although it is argued that the instability in the Fermi surface, causing the separation of hole and electron pockets (the nesting of Fermi surfaces), is the predominant mechanism, the underlying physics behind this phenomenon is a subject of debate. Similar behavior is also ascribed due to at least two other effects: (a) the distortion in the crystal structure leading to the Umklapp correction to the Landau free energy expression (for example, in CaFe$_{4}$As$_{3}$), and (b) step-like incommensurate peaks impersonating domain walls propagation of soft solitons that merge to the commensurate wave-vector via the first order phase transition (for instance, the lock-in transition in TaS$_{2}$).\cite{Manuel,Moncton}

\begin{figure}
\centering
\includegraphics[width=8.5 cm]{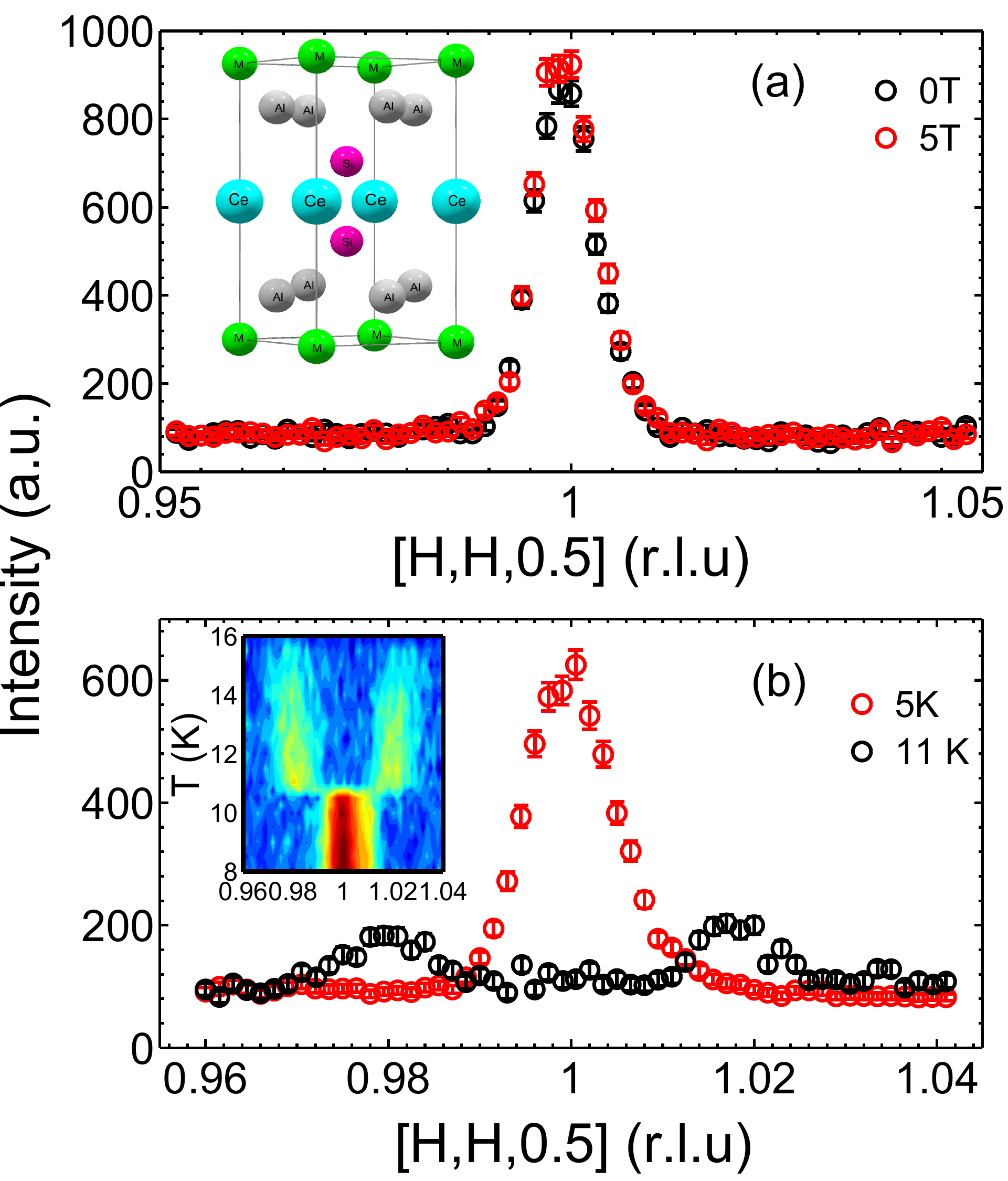} \vspace{-4mm}
\caption{(color online) Representative scans exhibiting commensurate and incommensurate magnetic reflections in Ce$X$Al$_{4}$Si$_{2}$. (a) Elastic scan along [HH0.5] reciprocal direction, depicting commensurate magnetic reflection, at $T$ = 5 K and at different field values in CeRhAl$_{4}$Si$_{2}$. The elastic data remains unaffected to magnetic field application up to $H$ = 10 T. Similar behavior is observed in $X$ = Ir. Inset shows the crystal structure. (b) Elastic scans along [HH0.5] direction, depicting incommensurate magnetic reflections, at $T$ = 11 K in CeIrAl4Si2. Similar behavior is observed in X=Rh. Inset shows the color map of incommensurate to commensurate transition as a function of temperature.
} \vspace{-4mm}
\end{figure}

We have performed detailed neutron scattering measurements on high quality single crystals of newly discoverd Kondo lattices Ce$X$Al$_{4}$Si$_{2}$, $X$ = Rh, Ir, to investigate the underlying magnetism as functions of temperature and magnetic field. We have found that both compounds undergo incommensurate-to-commensurate magnetic transition at $T$$_{c}$ $\simeq$ 9.2 K and 10.8 K in $X$ = Rh and Ir, respectively. The incommensurate magnetic order, which is field-independent and develops below $T$$_{IC}$ $\simeq$ 14 K in CeRhAl$_{4}$Si$_{2}$ (16 K in $X$ = Ir), is given by the temperature-dependent propagation wave-vector $\textbf{k}$ = (0.016,0.016,0.5) at $T$ = 10 K. The magnetic configuration in the incommensurate phase is best described by a spin density wave correlation of Ce-spins for 9 K $\leq$T $\leq$14 K (10.7 K $\leq$ T $\leq$15.5 K in $X$ = Ir), with Ce moments spatially fluctuating along the $z$- axis. As the temperature is decreased below T $\simeq$ 9 K (T $\leq$ 10 K in $X$ = Ir), the incommensurate (IC) peaks lock-in to the long-range antiferromagnetic order via the first order magnetic transition. The qualitative analysis of neutron data, combined with previous magnetic, electrical and heat capacity measurements on single crystal sample,\cite{Maurya} suggest that the lock-in magnetic transition arises due to the changes in the electronic properties of the system. The incommensurate magnetic structure at intermediate temperatures is related to the separation of electron and hole pockets in Ce$X$Al$_{4}$Si$_{2}$, as opposed to the lattice distortion causing the Umklapp correction to the free energy or the soliton propagation of domain walls.

Ce$X$Al$_{4}$Si$_{2}$, a dense Kondo lattice, crystallizes in the EuIrAl$_{4}$Si$_{2}$-type tetragonal lattice structure (space group P4/mmm) with lattice parameters of $a = b$ = 4.216 $\AA$ (4.221 $\AA$) and $c$ = 7.979 $\AA$ (7.949 $\AA$) in $X$ = Rh (Ir), as shown in the inset of Fig. 1a. Previous magnetic and heat capacity measurements on powder and single crystal specimens suggest strong anisotropic nature of magnetic susceptibilities, with large discrepancies in Curie-Weiss temperature, $\Theta$$_{CW}$, for field applications along different crystallographic directions and full ordered moment values in CeRhAl$_{4}$Si$_{2}$ and CeIrAl$_{4}$Si$_{2}$.\cite{Maurya,Ghimire,Ghimire2} In particular, for field application along [100] direction, $\Theta$$_{CW}$ and full moment values are found to be -155 K (-140 K) and 2.65 $\mu_B$ (2.62 $\mu_B$) in CeRhAl$_{4}$Si$_{2}$ ($X$ = Ir), respectively.\cite{Maurya} Magnetic and electrical measurements in applied field on single crystal CeRhAl$_{4}$Si$_{2}$ ($X$ = Ir) further reveal the onset of a spin-flop transition around $H$ $\simeq$ 5 T (6.5 T), which tends to disappear at T $\geq$ 10 K.\cite{Maurya}

Detailed neutron scattering measurements were performed on high quality single crystal samples of CeRhAl$_{4}$Si$_{2}$ and CeIrAl$_{4}$Si$_{2}$, with respective masses of 0.63 g and 0.17 g, on cold triple axis spectrometer SPINS at the NIST Center for Neutron Research and on thermal triple axis spectrometer TRIAX at the Missouri University Research Reactor. Single crystal samples were synthesized using the flux growth method and the high quality of the samples were verified using X-ray diffraction measurements.\cite{Maurya} Small samples with flat geometry reduces neutron absorption cross-section, thus help us in the quantitative analysis of the neutron data. For SPINS measurements, the single crystal sample was mounted at the end of a 1 K stick and cooled in $^{4}$$He$ environment in a 10 T magnet. For elastic measurements, the collimator settings were M-80$^{'}$-Be filter-Sample-Be filter-80$^{'}$-5 blades flat analyzer-detector. The measurements on TRIAX were performed using a closed cycle refrigerator with the base temperature of $\simeq$ 5 K. The collimator settings for TRIAX experiment were as follows: PG filter-M-60$^{'}$-Sample-PG filter-60$^{'}$-flat analyzer-detector. Single crystal samples were aligned in [HHL] plane of the reciprocal space, such that the applied field direction was along (-110).

\begin{figure}
\centering
\includegraphics[width=8.6 cm]{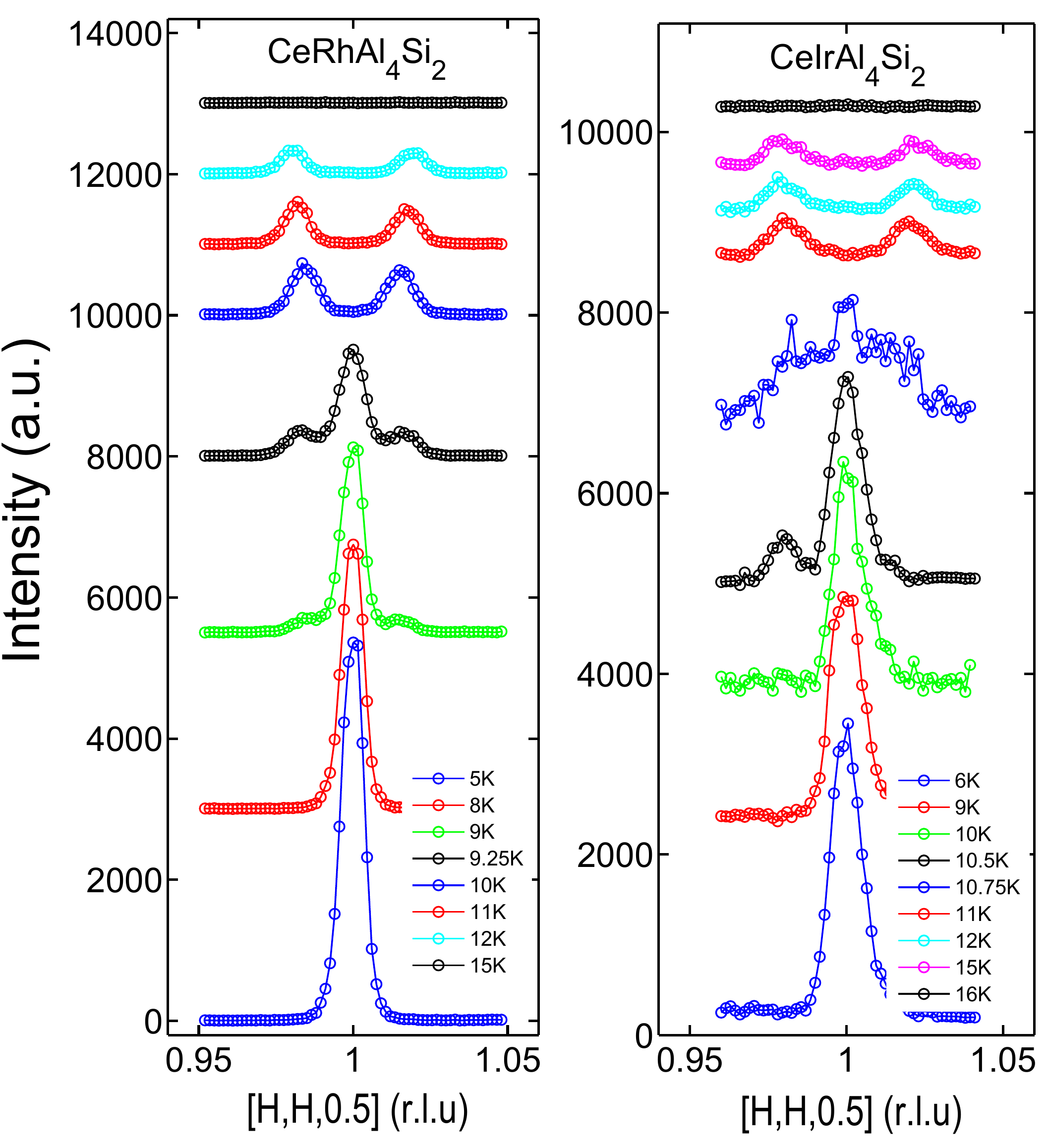} \vspace{-4mm}
\caption{(color online) Incommensurate to commensurate lock-in magnetic transition. (a) Elastic scans along [HH0.5] direction at different temperatures, exhibiting incommensurate to commensurate magnetic transition (T$_{c}$ $\simeq$ 9.2 K) as a function of temperature in CeRhAl$_{4}$Si$_{2}$. Experimental data are well described by a Gaussian lineshape. (b) Similar behavior is observed in $X$= Ir, albeit the transition happens at a slightly higher temperature (T$_{c}$ $\simeq$ 10.8 K).
} \vspace{-4mm}
\end{figure}

Previous neutron scattering measurements on powder Ce$X$Al$_{4}$Si$_{2}$, $X$ = Rh, Ir, were not conclusive enough to identify the magnetic correlation at intermediate temperature,\cite{Ghimire} as inferred from the susceptibility and heat capacity measurements.\cite{Maurya} Unlike the powder sample, single crystal specimen allows for a much more detailed investigation of structural and magnetic properties. As shown in representative Fig. 1b, the incommensurate magnetic pattern, indicating a different magnetic structure than the collinear antiferromagnetic configuration at low temperature, develops at intermediate temperature. Since both materials crystallize in the same tetragonal structure with same ligands coordination, the observation of incommensurate magnetic pattern in both systems, albeit with different onset temperatures (as discussed below), is not a surprise.

Next, we have performed measurements to study the evolution of spin correlation as a function of temperature. Representative scans along the reciprocal direction [HH0.5] at different temperatures in both compounds are plotted in Fig. 2a and 2b. As the measurement temperature is reduced below 14 K (16 K in $X$ = Ir), a pair of temperature-dependent incommensurate (IC) magnetic peaks develop with the propagation wave vector of $\textbf{$\textbf{k}$}$$_{IC}$ = (0.016,0.016,0.5) at $T$ $\simeq$ 10 K. The position of IC peak, with respect to the nearest nuclear peak, is described by the wave vector: $\textbf{q = Q $\pm$k}$, where $\textbf{Q}$ represents the nuclear peak position. As the temperature is further reduced, the incommensurate peaks first get stronger before gradually diminishing to the background level at $T$ $\simeq$ 9.2 K (10.8 K in $X$ = Ir). Around the same temperature, new magnetic peaks with the commensurate propagation vector $\textbf{k}$$_{C}$ = (0,0,0.5) develop in both compounds. The commensurate magnetic peaks become stronger as temperature is reduced to T $\rightarrow$ 0 K. The overall behavior is described in Fig. 3a and 3b, depicting the temperature dependence of magnetic peak intensity in both commensurate and incommensurate phases. Unlike in the incommensurate phase where a dome-shaped regime with the maximum intensity around T $\simeq$ 10 K (11 K in $X$ = Ir) occurs in temperature, the sharp temperature-dependence of the order parameter in the commensurate phase (especially in $X$ = Ir compound) suggests a first order magnetic transition. However, elastic neutron data did not exhibit any magnetic field dependence, as shown in Fig. 1b.

\begin{figure}
\centering
\includegraphics[width=8.8 cm]{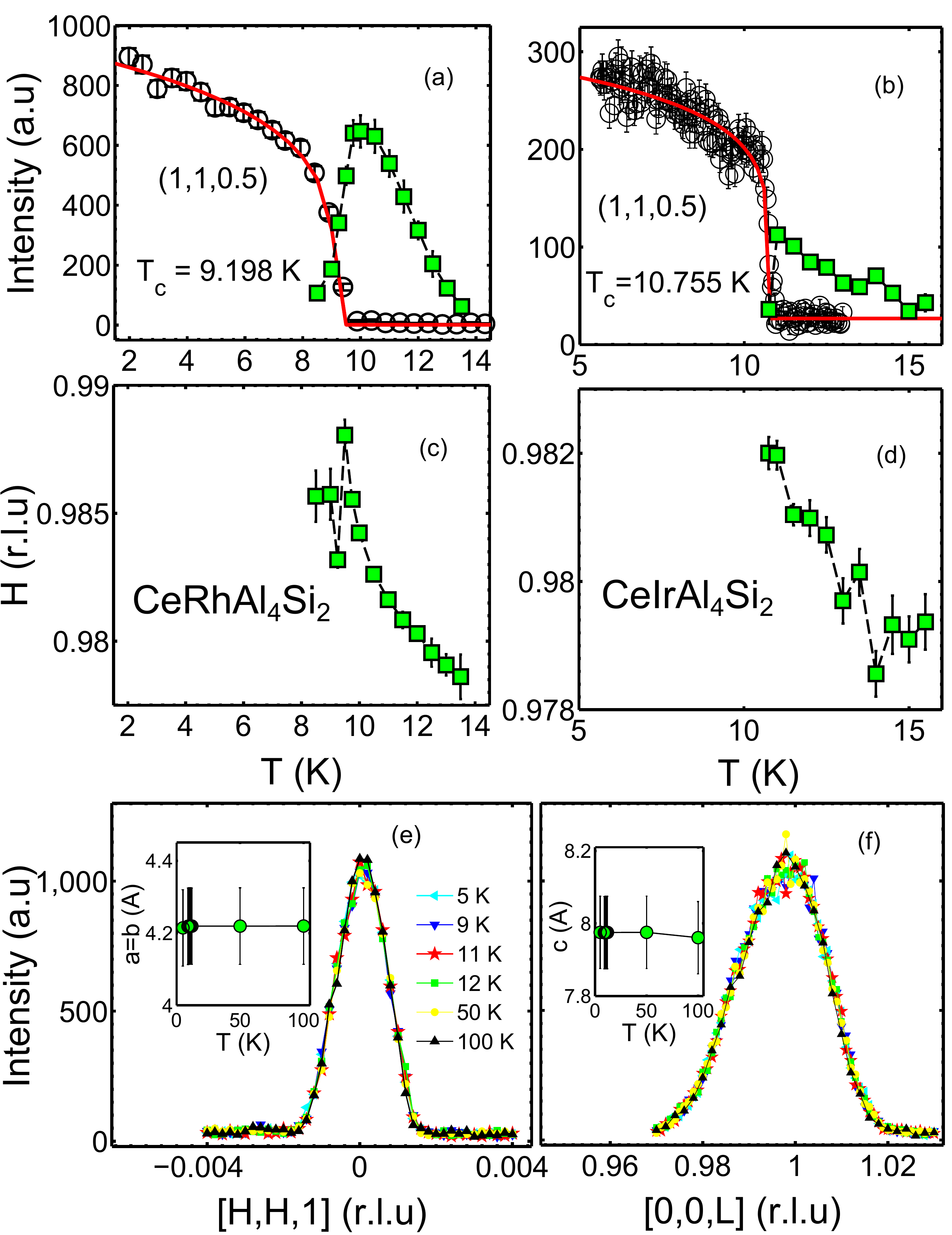} \vspace{-4mm}
\caption{(color online) Commensurate and incommensurate magnetic order parameter and movement of IC peaks. (a) and (b) Magnetic order parameters in both commensurate and incommensurate (IC) phases as a function of temperature in CeRhAl$_{4}$Si$_{2}$ (Fig. a) and CeIrAl$_{4}$Si$_{2}$ (Fig. b). While the IC peaks form a dome-type structure in temperature, the transition to the commensurate phase is mostly first order in nature. (c) and (d) Center of IC peak as a function of temperature, nominally at $\textbf{q}$$_{IC}$ = (0.98,0.98,0.5). The IC peak moves towards the commensurate wave-vector as the measurement temperature is reduced.  (e) and (f) Representative scans across structural peaks as a function of temperature in CeRhAl$_{4}$Si$_{2}$ (Fig. e) and CeIrAl$_{4}$Si$_{2}$ (Fig. f). Clearly, no change in the position or the peak intensity of nuclear peaks are observed as the temperature is reduced through the lock-in magnetic transition, ruling out the effect of crystal distortion in the occurrence of the lock-in magnetic transition.
} \vspace{-4mm}
\end{figure}

Magnetic peaks in Ce$X$Al$_{4}$Si$_{2}$, Fig. 2, are best described by a Gaussian lineshape of width limited by the instrument resolution. It indicates the presence of long-range magnetic order in the system. More quantitative information is derived from the elastic neutron data that help us further understand the underlying magnetism in Ce$X$Al$_{4}$Si$_{2}$. In Fig. 3c and 3d, we have plotted the temperature dependence of the incommensurate peak, nominally at $\textbf{q}$$_{IC}$ = (0.98,0.98,0.5). As the measurement temperature is reduced, IC peak gradually moves towards the commensurate wave-vector. The spectral weight shifts from IC peak to the commensurate peak in a very narrow temperature range around $T$$\simeq$9 K ($\simeq$ 10.7 K in $X$ = Ir), suggesting the lock-in magnetic transition in the system (see the inset of Fig. 1b and Fig. 2). In principle, this behavior can arise due to any of the three reasons: lattice distortion, soft soliton creation or the changes in the electronic properties due to the Fermi surface reconstruction. According to the Landau-Lifshitz expression for the free energy, the magnetic order parameter $\eta$ can transform from an incommensurate to a commensurate phase due to the Umklapp correction to the total energy, given by:\cite{Crawley,Manuel}
\begin{eqnarray}
{G}&=& {G_0}+{\frac{1}{2}}{a}{(T-T_c)}{\eta}^2+ {u}{\eta}^4+{V}{\eta}^{p}{cos{p}{\varphi}}   
\end{eqnarray},
where G is the free energy, $T_c$ is the transition temperature and $\varphi$ is the phase of the distorted wave, which can take any of the $p$-values. Minimization of the above expression for $\eta$ yields three values of $\varphi$, corresponding to three different lattice structures associated to the first order transition to the commnesurate magnetic phase.\cite{Crawley} The Umklapp correction is more prominent in the case of significant lattice distortion in the system, primarily causing the crystal symmetry group transformation at low temperature. Measurements were performed to determine the applicability of this possibility. Representative scans across structural peaks and the lattice parameters in Ce$X$Al$_{4}$Si$_{2}$ are plotted in Fig. 3e and 3f. Clearly, the tetragonal lattice structure remains intact throughout the measurement; hence, rules out the possibility of the Umklapp correction causing an incommensurate-to-commensurate magnetic phase transition in Ce$X$Al$_{4}$Si$_{2}$. Second possibility involves the creation of soft solitons at higher temperature that merge to the commensurate wave-vector via the first order transition as the measurement temperature is reduced.\cite{Crawley,Moncton} Soft solitons are accompanied by a step-like function of incommensurate (IC) reflections of higher harmonics. Also, the IC peaks get closer to each other as temperature is reduced. Ce$X$Al$_{4}$Si$_{2}$ exhibits at least two characteristics of solitons: a first order-type magnetic transition to the commensurate phase (more prominently in $X$ = Ir) and the temperature-dependent movement of IC peaks towards the commensurate wave-vector. However, no evidence of the step-like function of higher harmonic IC peaks were observed in neutron scattering measurements. More research is needed to further verify or, completely rule out the possibility of soliton propagation in this dense Kondo lattice system. Under present circumstances, it is imperative to consider that the changes in the electronic properties, involving the separation of electron and hole pockets at intermediate temperatures, causes the lock-in magnetic transition in Ce$X$Al$_{4}$Si$_{2}$. In a more recent report, it was shown that the compound $RM$Al$_{4}$Si$_{2}$, where $R$ = La, Ce and $M$ = Rh, Ir, Pt, have quasi-2D electronic structure with electron and hole pockets separated by a similar wave-vector.\cite{Ghimire2} This is in agreement with our qualitative analysis.

\begin{figure}
\centering
\includegraphics[width=8.9 cm]{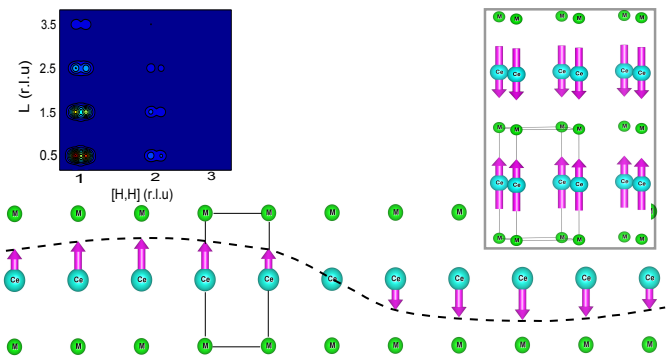} \vspace{-4mm}
\caption{(color online) Top left panel manifests the simulated magnetic pattern in the incommensurate phase, which is consistent with the experimental data. Top right inset shows the spin correlation of Ce-ions in the commensurate phase. While the commensurate phase is described by the antiferromagnetic correlation of Ce-ions along the $z$-axis, the incommensurate structure (as shown in lower panel) is best described by a spin density wave with Ce-spins spatially fluctuating along the $z$-axis.
} \vspace{-4mm}
\end{figure}

Finally, we discuss the nature of long range magnetic correlation in Ce$X$Al$_{4}$Si$_{2}$. The incommensurate magnetic reflections are usually associated to the long-range magnetic order of a density wave or the square wave pattern. Measurements were performed to higher order Brillouin zones (BZ) at two temperatures, $T$ = 5 K and 10 K, to understand the nature of magnetic correlations. At low temperature, magnetic peak intensities across the extended BZ is best described by an antiferromagnetic spin correlation with Ce-spins pointing along the $z$-axis, see the inset of Fig. 4. The ordered moment of correlated Ce-ions is found to be 1.12(0.17) $\mu_B$ and 0.89(0.16) $\mu_B$ in $X$ = Rh and Ir, respectively. This is also consistent with a previous report of neutron scattering measurements on powder Ce$X$Al$_{4}$Si$_{2}$.\cite{Ghimire} In order to determine the spin correlation associated to IC peaks at relatively higher temperature ($T$$\simeq$10 K), numerical modeling of the experimental data was performed using the following expression for the ordered moment $\textbf{S}$$_{ij}$:\cite{Bao,Berger}

\begin{eqnarray}
{\textbf{S}}_{\textbf{ij}}&=& {{\textbf{A}}_{i}{cos({\textbf{k.r}_{\textbf{j}}}+{\psi}_{j})} + {\textbf{B}}_{i}{sin({\textbf{k.r}_{\textbf{j}}}+{\psi}_{j})}}
\end{eqnarray},

where $\textbf{S}$$_{ij}$ is the moment of the $i$th ion in the $j$th unit cell and $\textbf{k}$ is the propagation wave vector of the spin density wave.\cite{Berger} The numerical modeling also involved averaging the magnetic structure factor over four equally populated domains. Magnetic structure in the IC phase is best described by a spin density wave (SDW) configuration of propagation vector $\textbf{k}$ = (0.016,0.016,0.5). The spin correlation of Ce spins, spatially fluctuating along the $z$-axis, is shown in Fig. 4. The ordered moment associated to the SDW configuration is found to be 0.65(0.2) $\mu_B$ and 0.45(0.16) $\mu_B$ in $X$ = Rh and Ir, respectively. The ordered moment values, in both the commensurate and incommensurate phases, are much smaller than full moment values in respective compounds. Since Ce$X$Al$_{4}$Si$_{2}$ is a dense Kondo lattice material, the smaller value of ordered moment possibly illustrates strong Kondo screening of the localized moment by surrounding conduction electrons.

In summary, we have performed detailed experimental investigation of the underlying magnetism and associated lock-in magnetic transition in single crystals Ce$X$Al$_{4}$Si$_{2}$. Both compounds exhibit sharp magnetic transition from an incommensurate phase, at intermediate temperatures, to the commensurate phase at low temperatures. The spin structures in commensurate and incommensurate phases are manifested by long range antiferromagnetic and spin density wave configurations of correlated Ce-ions, respectively. The qualitative analysis of the experimental data, in conjuction with a recently published report,\cite{Ghimire2} suggests that the incommensurate phase can be arising due to the Fermi surfaces nesting, which causes the separation of electron and hole pockets at intermediate temperatures. Further theoretical works involving the calculation of the propagation wave vector in a nested Fermi surface, using Linear Muffin Tin Orbital (LMTO) method or other analytical techniques, are very desirable. Previously, similar theoretical methods were employed to establish the correlation between the Fermi surfaces nesting and the incommensurate magnetic structure in heavy fermion compounds, such as CeCu$_{2}$Si$_{2}$ and CeCu$_{2}$Ge$_{2}$.\cite{Stockert,Zwicknagl} Also, detailed inelastic neutron scattering measurements on bigger samples will be helpful in further understanding the role of possible soliton propagation in these newly discovered Kondo lattices.\cite{Crawley}  

Authors acknowledge the support provided by the Department of Commerce facility NIST Center for Neutron Research. Authors are thankful to J. Leo and A. Ye for help with the neutron scattering experiments.

\clearpage

\end{document}